# ESTIMATION OF OPTIMIZED ENERGY AND LATENCY CONSTRAINT FOR TASK ALLOCATION IN 3D NETWORK ON CHIP

Vaibhav Jha[1], Mohit Jha[2] and G K Sharma[1]

[1]Department of Computer Science & Engineering, ABV- Indian Institute of Information Technology and Management, Gwalior, Madhya Pradesh 474015
vaibhavjha1987@yahoo.com

[2]Department of Electrical Engineering, Jabalpur Engineering College, Jabalpur, Madhya Pradesh 482011
mohitjha_1989@yahoo.com

*ABSTRACT*

*In Network on Chip (NoC) rooted system, energy consumption is affected by task scheduling and allocation schemes which affect the performance of the system. In this paper we test the pre-existing proposed algorithms and introduced a new energy skilled algorithm for 3D NoC architecture. An efficient dynamic and cluster approaches are proposed along with the optimization using bio-inspired algorithm. The proposed algorithm has been implemented and evaluated on randomly generated benchmark and real life application such as MMS, Telecom and VOPD. The algorithm has also been tested with the E3S benchmark and has been compared with the existing mapping algorithm spiral and crinkle and has shown better reduction in the communication energy consumption and shows improvement in the performance of the system. On performing experimental analysis of proposed algorithm results shows that average reduction in energy consumption is 49%, reduction in communication cost is 48% and average latency is 34%. Cluster based approach is mapped onto NoC using Dynamic Diagonal Mapping (DDMap), Crinkle and Spiral algorithms and found DDmap provides improved result. On analysis and comparison of mapping of cluster using DDmap approach the average energy reduction is 14% and 9% with crinkle and spiral.*

## 1. INTRODUCTION

The scaling of microchip technologies has resulted into large scale Systems-on-Chip (SoC), thus it has now become important to consider interconnection system between the chips. The International Technology Road-map for Semiconductors depicts the on-chip communication issues as the limiting factors for performance and power consumption in current and next generation SoCs [1] [2] [3]. Thus Network on chip has not only come up with an alternative for the SoC, it has also solved the problem faced in the traditional bus based architecture and is an efficient approach towards optimal designs. Although various works has been done in the optimization of the design and the major area where the design need to be focused are Topology, Scheduling, Mapping, and Routing [2]. Each area plays an important role in delivering better performance of the system, but in this paper emphasis is been given on the scheduling and mapping of the IP core onto the 3d architecture.

Application scheduling and core mapping have a significant impact on the overall performance & cost of network on chip (NoC). The application is described as a set of concurrent task that have been assigned and scheduled to NoC [4]. An application is described through its Communication Task Graph as shown in Figure 1. A scheduling algorithm is then used to assign application tasks (threads) to available IP cores and to specify their order of execution. After the scheduling step, the Application Characterization Graph is obtained. After that mapping algorithm is used to assign tasks to processing elements and then perform a packet routing function.

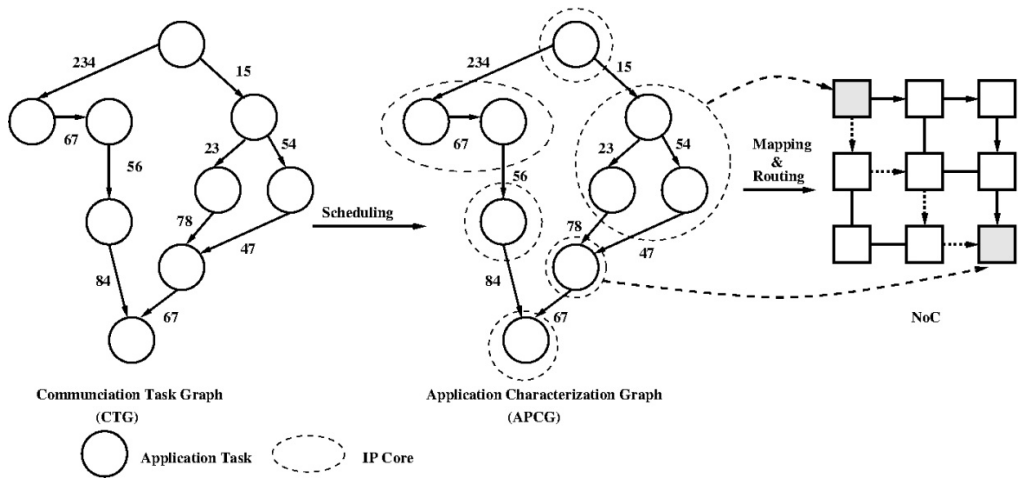

Figure 1. Application Scheduling, Mapping and routing problem

We observe that both scheduling and mapping algorithms for Networks on Chip have similar objectives. Increasing the performance and decreasing the energy consumption of a NoC, for a particular application two optimizations typically made by such algorithms. Ideally, both scheduling and mapping problems should be treated together. In other words "scheduling" means mapping the application tasks onto the available IP cores, and "mapping" means mapping the IP cores onto the available NoC nodes. Therefore, both scheduling and mapping problems deal with application mapping onto a Network-on-Chip. Application scheduling is to assign each task in a task graph to different cores and decide the sequence of their execution (called execution table) if two or more task are scheduled on the same core.

By mapping of the IP core we mean, assigning the task in the form of the characterization graph to the given architecture following the design constraint such as area, latency, communication energy and communication cost which should be minimum. As today's application are becoming much more complex their average communication bandwidth are in Gbps and as technology is scaling down, in future a single application would be running on single core thus the bandwidth requirement of link would increase, therefore attention is also given onto its minimization. Figure 2 shows the design flow of scheduling and mapping of the tasks onto 2D mesh architecture. Assigning the given application onto the given architecture is very important from energy point of view, sometime the topology also matters more in optimizing the design [5]. Topology helps in determining latency, traffic pattern, routing algorithm, flow control, and makes huge impact on design cost, power, and performance. Mapping of the IP core onto NoC tile could be done by either assigning a single core onto each tile or by assigning multiple IP core on each of the tile i.e., Scheduling. Each of the mapping procedure has its own merits and demerits. In this we are presenting a similar approach of allocating IP cores on each NoC processing elements.

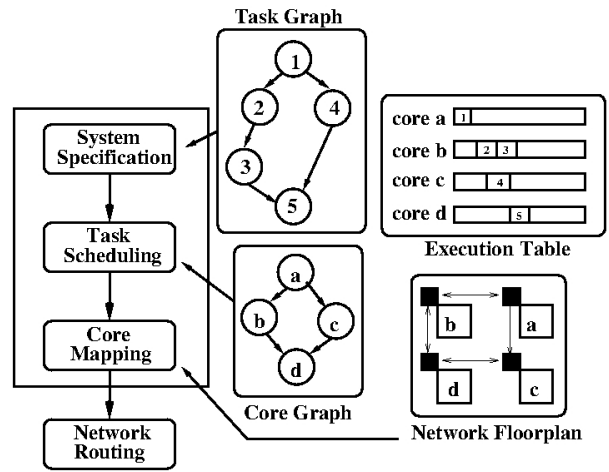

Figure 2. Network on Chip Design Flow.

In addition [6], include a automated design technique based on particle swarm optimization that solve the problem of routing decision with the objective of minimum communication energy consumption. PSO is a technique which randomly places an IP core and tries to minimize the objective function based on swarm intelligence.

## 2. RELATED WORK

Various mapping algorithm has been discussed by the researchers in their literature. *Branch-and-Bound* [7], *Binary Particle Swarm Optimization* (BPSO) [8], based on latency and energy aware, *Diagonal map* [9], are few mapping algorithm which is for 2D NoC architecture but they are also applicable on 3D architecture and shows better results compared to 2D. Task priority based [10], multiple $v_{dd}$ [11] [12] and thermal aware mapping [13], are few work done by the researchers in the field of 3D NoC.

Energy and performance aware mapping using Branch and Bound (PBB) [7] maps the IP core on the basis of the PBB technique in which mapping complexity increases exponentially as the number of the core increases. Another concept of mapping in which the author has focused on latency and energy using the BPSO [8] has proposed mapping as the NP-hard problem and thus the heuristic approach is needed for finding the solution. Author has compared result with the Genetic algorithm and found BPSO optimal. In D-map [9] procedure, mapping is based on the concept that IP core which is communicating maximum with the rest of the core should be placed on those NoC tile which has greater number of the neighboring tile attached to it thus in 2D architecture diagonal tiles are having maximum number of the neighboring tiles.

In mapping procedure using *multiple $v_{dd}$* Kostas and Dimitrios, [11] [12] has claimed energy of the system could be saved either applying better mapping or the routing procedure or by supplying less voltage to the system and making the design itself optimal. As not each of the router function every time so author divided the whole of the architecture into two layer and each of the layer functioning at different voltage of $v_{dd_{high}}$ and $v_{dd_{low}}$. IP core with greater communication volume are placed onto layer with $v_{dd_{high}}$ and those with low in $v_{dd_{low}}$ when they need to communicate they do with level converter. In his continuing paper [11] author has compared results with different benchmark of MWD, MPEG, VOPD, MMS etc.

In [10] author has proposed two different method of mapping procedure on 3D NoC design *Crinkle* and *Spiral* and task are organized on the task priority list which is based on the basis of the maximum communication volume or maximum out-degree. In [13], author has targeted communication aware used the *Genetic algorithm* for finding out the optimal mapping solution.

In [8], author proposed a heuristic mapping algorithm based on chaotic discrete particle swarm optimization technique. The algorithm resolves the mapping problem to optimize the delay and energy consumption and showing better results than genetic algorithm. In [6], author proposed a routing technique, based on the Hybrid Particle Swarm Optimization (PSO) Algorithm is applied on the 2D-Mesh NoC platform to balance the link load and this algorithm is combined with genetic algorithm, and shown a better results. In [14] author has proposed optimal multi-objective mapping algorithm for NoC architecture.

In [15] Energy aware scheduling which is based on the idea of slack budgeting is discussed, which allocates more slack to those tasks whose mapping onto PE has larger impact on energy consumption and performance. It is carried out in three phase: 1. Budget slack allocation for each task 2. Level based scheduling 3. Search and Repair

Energy aware mapping and scheduling EAMS [16] is given which first partitions the task into two groups: critical and non-critical tasks. The tasks that have closer deadline and there is a probability of deadline failing are called critical tasks, while the others are called non-critical tasks. It first schedule and map critical tasks then after non critical tasks.

Cluster based application mapping for NoC is discussed in [17]. In this method author said, decompose the mapping problem into several sub-problems and solve each sub-problem separately. In other words, first decompose the mesh into sub-meshes and partition the application graph into smaller sized clusters. After that applied ILP-based mapping method for each cluster-sub-mesh pairs. Clustering algorithm [18] [19] map the tasks in a given DAG (directed acyclic graph) to an unlimited number of processors. Each iteration refines the clustering by merging some clusters until the number of clusters is equal to the number of processors.

Thermal aware task scheduling for 3d multiprocessor is provided in [20] [21] [22]. In this paper, a heuristic OS-level technique is proposed that performs thermal-aware task scheduling on a 3D chip multiprocessor (CMP). The proposed technique aims to improve task performance by keeping the temperature below the threshold to reduce the amount of dynamic thermal management (DTM). They used dynamic thermal management (DTM) techniques such as dynamic voltage and frequency scaling (DVFS).

## 3. DEFINITIONS

Definition 1: A Communication Task Graph (CTG) G = G (T,C) is a directed acyclic graph, where each vertex represents a computational module of the application referred to as a task $t_i \epsilon T$. Each $t_i$ has the following properties:
- An array $R^i$, where the $j^{th}$ element $r^i_j \epsilon R^i$ gives the execution time of task $t_i$ if it is executed on $j^{th}$ PE in the architecture.
- An array $E^i$, where the $j^{th}$ element $e^i_j$ gives the energy consumption of task $t_i$ if it is executed on $j^{th}$ PE in the architecture.
- A deadline $d(t_i)$ which represents the time $t_i$ has to finish. If the designer does not specify a deadline for task $t_i$, then $d(t_i)$ is taken equal to infinity.

Definition 2: An application characterization graph (APCG) G = (C,A) is a directed graph, where each vertex $c_i$ represents selected IP/core, and each directed arc $a_{i,j}$ characterizes the communication from $c_i$ to $c_j$. Each $a_{i,j}$ has application specific information such as:

- $v(a_{i,j})$ is the arc volume from vertex $c_i$ to $c_j$, i.e. the communication volume (bits) from $c_i$ to $c_j$.
- $b(a_{i,j})$ is the arc bandwidth requirement from vertex $c_i$ to $c_j$.

Definition 3: A NoC architecture can be uniquely described by the triple Arch( *T(R, Ch, L), $P_R$ , $\Omega(C)$* ), where:

1) T(R,Ch,L) is directed graph describing the topology. The routers (R), channels (Ch) and the Layers (L) in the network have following attributes:
    a) ∀ *(ch) ϵ Ch*, *w(ch)* gives the bandwidth of the network channels.
    b) ∀ *(r) ϵ R, I(ch, r)* gives the buffer size(depth) of the *channel ch*, located at *router r*.
    c) ∀ *(r) ϵ R, Pos(r)* specifies the position of *router r* in the floor plan.
    d) ∀ *(l) ϵ L, Layer(l)* specifies the layer of topology.
2) $P_R$(r,s,d) describes the communication paradigm adopted in the network.
    - s,d,r ϵ R, n ⊂ R defines the routing policy at *router r* for all packets with *source s* and *destination d*.
3) $\Omega : C \to R$, maps each core $c_i$ ϵ C to a router. For direct topologies each router is connected to a core, while in indirect topologies some routers are connected only to other routers.

## 4. PROBLEM FORMULATION

As we are aiming for the minimization of the total communication energy and the total communication energy ($E_{total}$) depends on *Latency, number of hops count, links energy, and switch energy* [7] [14], therefore minimization of each of these factor will result into reduction of global $E_{total}$. Thus our problem has been formulated as:

**Given** an APCG *G(C,A)* and a network topology *T(R, Ch, L)*;
**Find** a mapping function $\Omega$ : C → R which maps each core $c_i$ ϵ C in the APCG to a router *r* ϵ R in *T(R, Ch, L)* so that we get :

$$min \{ \sum v(a_{i,j}) \times e(r_{map(c_i),map(c_j)}) \}$$

**such that:**
$$\forall c_i \in C, map(c_i) \in T$$
$$\forall c_i \neq c_j \in C, map(c_i) \neq map(c_j)$$

where $e(r_{i,j})$ is the average energy consumption of sending 1-bit of data from tile $t_i$ to $t_j$ on which core $c_i$ and $c_j$ are mapped respectively.

Various energy consumption model has been discussed in [7][2], the 1 bit energy consumption in the NoC architecture as proposed in [7] which is calculated as:

$$E_{Bit}^{t_i,t_j} = n_{hops} \times E_{S_{Bit}} + (n_{hops} - 1) \times E_{L_{Bit}} \qquad (1)$$

where, $E_{L_{Bit}}$ is the per bit link energy, $E_{S_{Bit}}$ is per bit switch energy, $E_{Bit}^{t_i,t_j}$ is the total energy when one bit is moved from tile $T_i$ to tile $T_j$, $n_{hops}$ represent the total number of the hops.

Based on below proposed algorithm performance of the system will increased as our approach reduces the number of hops between source and destination. The performance of the system is evaluated on the basis of total communication cost which is calculated as

$$Cost = \sum_{\forall j=1,2,3\ldots|V|, i \neq j} \{ b_{a_{i,j}} \times n_{hops}(i,j) \}$$

**Latency:** In a network, latency is a synonym for delay, is an expression of how much time it takes for a packet of data to get from one designated point to another and so that average latency is calculated as:

$$Latency_{avg} = \frac{\sum_{\forall j=1,2,3\ldots|V|, i \neq j} \{ n_{hops} \times Comm.Volume \times \rho \}}{\eta}$$

where, $\rho$ = constant related to delay
$\eta$ = Total number of times transfer occurred of communication volume between source and destination.

## 5. PROPOSED ALGORITHM

Assignment of an IP core onto the given architecture is called a *mapping*. When we map an IP core onto the given architecture we always try to place those core which is communicating maximum($core_{i,j}^{max}$) closer to each other, limiting the number of hops travelled by data between two related cores, thus the $E_{total}$ gets reduced. As in 2D architecture diagonal tiles(tiles with degree 4) are having greater number of communicating links, so placing $core_i^{max}$ onto these tile and placing rest of unmapped core w.r.t this mapped core onto neighboring tile would reduce the number of hop count thus $E_{total}$ would minimize [9].

Similar approach when carried out with the 3D NoC in which the topology is 2D layered architecture shown in Figure 3, the diagonal tiles of each layer has four adjacent links and one *uplinks* and *downlinks* each. The mapping of highest out-degree core onto diagonal tiles gives the flexibility of assigning highest communicating unmapped cores much closer to mapped cores. This helps to minimize the number of hops travelled by data between two communicating core, which is our aim to minimize the total energy consumption. Question arises how to find the IP core and arrange them such that it could be mapped in regular fashion. In this paper we have used the concept of maximum out degree. We find those cores which is having maximum out degree and order them in an array of descending order, if there are two or more IP cores which is having same number of the out-degree then we differentiate them on the basis of the *Rank* which is calculated as:

$$Ranking\ (Core_i) = \sum_{\forall j=1,2,3\ldots|V|, i \neq j} (comm_{i,j} + comm_{j,i})$$

where $comm_{i,j}$ represents the communication volume(in bits) transferred from $core_i$ to $core_j$. The core to be mapped next is the unmapped core which communicates maximum with the mapped core.

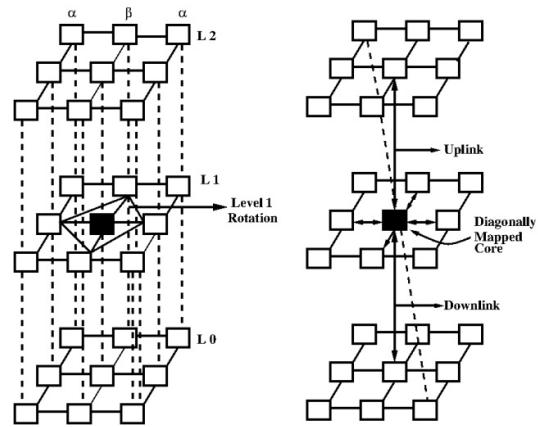

Figure 3. Dynamic Diagonal Mapping Technique.

After arranging the IP cores on the basis of the out-degree and rank, the core with maximum out-degree is mapped onto the diagonal tiles. Each layer in the architecture is having only one tile which comes in the list of the diagonal tiles to which these highest out-degree IP core will be mapped. While mapping we leave the diagonal end tiles as it is having at most only three tile to which it can communicate. After mapping the core onto the diagonal tile the position of the next unmapped core is found on the basis of the lozenge shape, in which we apply two rotations namely $\alpha$ and $\beta$ rotation. $\alpha$ Rotation works out in clockwise direction for odd numbered column and $\beta$ in anti-clockwise direction for even numbered column discussed in literature [9]. While mapping the IP core onto the each layer we apply this rotation for intra-layer tiles only for finding the position of the next empty tile, but during the mapping process their is possibility that position of the next empty tile could not be there in the same layer so we need to change the layer. While applying the rotation we need to change the level of the rotation, the maximum number of the level which can be changed in each rotation is *2×(n-1)*, where *n* is the dimension of the architecture(*n×n×n*).

While changing the layer for finding the position of the next empty tile, the position in next layer from where the search will start remains the same as that of the mapped core in the previous layer. Same as we applied α or β rotation for finding the position of the next empty tile. The variable layer increases if mapped core is in layer 0, it decreases if mapped core is in layer n-1 or else it may increase or decrease if it's in between. In calculation of the energy, latency is also important which is dependent on the routing procedure. In our work we have used the XYZ routing algorithm for finding out the number of the hops count.

In this way we can able to schedule more than one IP core to the processing element, when all processing element are exhaust but if still we have IP cores, so we can still able to schedule IP core over the processing element (PE) i.e., we can start again with the same method for the scheduling of remaining IP cores and we are not just limiting to the number of IP core should be equal to the number of processing element. So, our idea first allocate one IP to all available processing element and then if we have still remaining IP core so we start with allocating two IP core to the PE and three, four and so on.

- Firstly, this algorithm utilizes all the resources, when number of IP core is more than the available resources. If IP are less than the available resources then grab resources equal to IP cores.
- Then, after scheduled two or more IP cores to the PE until IP vanishes according to the algorithm procedure discussed in Algorithm 1.

In calculation of the energy, latency is also important which is dependent on the routing procedure. In our work we have used the *XYZ* routing algorithm for finding out the number of the hops count.

Following the Algorithm 1 the number of hops count between the two related cores is reduced, which in turn reduces the total communication energy consumption given in Equation 1.

**Algorithm 1** Mapping Algorithm for 3D NoC
--------------------------------------------------------------------------------------------------------------------
**Input:** APCG *G(C,A)* and *(n×n×n)* 3D NoC Topolcgy *T(R, Ch, L)*.
**Output:** Array *M[ ]* containing corresponding tile numbers for $c_i \in C$.
1: **Initialize:** *Mapped[ ] = -1;UnMapped[ ] = $c_i \in C$*
2: **Mapping()** {
3: *Sort ($c_i \in C$)*
4: *Store in OD[ ]. { in descending order of outdegree & Ranking().}*
5: **for** *i = 0 to n − 3* **do**
6:      *OD[ i ].mapto() → ($n^2 + n + 1$ ) × (i + 1);*
7:      *Mapped[ i ] = OD[i];*
8:      *Remove OD[ i ] from UnMapped[ ];*
9: **end for**
10: **while** *UnMapped[ ] ≠ empty* **do**
11:      for all *$c_i \in$ UnMapped[ ]*, select *$c_i$* where:
12:      *max comm(i,j) { where i ∈ UnMapped[ ] , j ∈ Mapped[ ];}*
13:      *positionof(j); //(row, column, layer) of $c_j$*
14:      *colno = (j.mapto())%n*
15:      **if** *(colno% 2 ≠ 0)* **then**
16:           **while** *(flag = 0)* **do**
17:                *alpharotation()*
18:                **if** *empty tile found* **then**
19:                     *return $t_k$, set flag = 1;*
20:                **else**
21:                     *layer+ + || layer- - || (both)*
22:                     *alpharotaion()*
23:                **end if**
24:           **end while**
25:      **else**
26:           **while** *flag = 0* **do**
27:                *betarotation()*
28:                **if** *empty tile found* **then**
29:                     *return $t_k$, set flag = 1*
30:                **else**
31:                     *layer+ + || layer- - || (both)*
32:                     *betarotation()*
33:                **end if**
34:           **end while**{returns empty tile($t_k$) using Lozenge shape path.}
35:           *OD[ i ].mapto() → $t_k$*
36:           *Mapped[i] = OD[ i ]*
37:           *Remove OD[ i ] from UnMapped[ ]*
38:      **end if**
39: **end while**
40: calculate_energy() {calculates energy for the mapping generated.}
--------------------------------------------------------------------------------------------------------------------

*Energy Aware Cluster Based Scheduling for 3D NoC:*

Scheduling tries to keep communication local (on the same processor) whenever possible or, if not feasible, tries to minimize the distance between two communication partners.

Each task in CTG transfers communication volume with its communicating partner node to a processing element to which it is mapped.

A task is the smallest unit for scheduling. The order in which the tasks of a given application can be executed is determined by data dependencies, which are captured in our approach by a directed graph called the task graph. Our scheme work sin three steps 1. Scheduling 2. Mapping 3. Packet routing.

Our task scheduling algorithm takes the number of virtual processing elements P and the task graph (CTG) of the application as input. It determines to which virtual processing element each task should be assigned. Our goal in this step is to exploit the parallelism in the application code as much as possible, and maximize data reuse within each processor by scheduling the tasks that share a large number of data blocks on the same virtual processing element. The goal of the task scheduling step is to cluster the tasks that share an enormous amount of data blocks among them into the same processor. In order to isolate task scheduling from processor mapping, we first schedule the tasks on to a set of virtual processing elements, which will be explained shortly. After that, in the processor mapping step, we map the virtual processing element onto physical processing element to improve inter processor data locality. That is, two virtual processing elements sharing a large amount of data are mapped to a pair of neighboring physical processors or PE's.

Finally, we determine the routings for the data packets, i.e., the set of links that are utilize during data accesses. Note that the cost for accessing a data block is determined by the distance between the processing element that issues the access and the node that contains the data block. Specifically, a longer distance to data (which is a measure of locality) means that the access request and the data block need to be transferred over a larger number of network links, and thus, incurs more energy consumption and longer delay (depending on the network switching mechanism employed). Our compiler based approach improves performance and reduces energy consumption by mapping the tasks and the data blocks into the chip multiprocessor nodes such that we can reduce the distances between the processors and the data blocks that are frequently accessed by these processors.

So, our motive is to minimize the communication energy between IP cores, by keep the communicated IP core local. Initially we select a first node and scheduled over virtual processing element and then travel a list of IP core partner of the scheduled IP core and scheduled the communicated neighbor partner with the scheduled one to maintain the control dependencies i.e., cluster with the parent node. If two or more neighbor partner comes so in that case choose the task which has maximum communication cost because our aim is to reduce overall communication energy consumption. This step is doing until we found a loop with the already scheduled IP core shown in Figure 4. After that in sequence choose the other unscheduled task and scheduled it over other virtual processing element and repeat the procedure until all task vanishes.

Here in this example we take up to only v10 because when the loop reach on node v10 and found its communicating partner is with already scheduled partner means already scheduled IP core has some dependency with the increasing loop node. So, we keep the looped IP core and discard the other. And the other nodes are scheduled over other IP cores. One point to note is here cycle communication, when such condition occur stop the searching of dependency neighbor partner and continue the scheduling of remaining IP cores to different processing element.

In Figure 5 for simplicity representation here we show the cluster of process. By using this approach we need only six processing element in place of sixteen and also this approach save the needed communication energy as compare to just map the IP core over processing element. Then after this, place these scheduled table clusters over NoC architecture in such a way to transfer packet from source to destination processing element or cluster number of hops become minimized, for that purpose we used a diagonal mapping algorithm of IP cores discussed above in dynamic scheduling, mapping is the part of scheduling in dynamic scheduling algorithm.

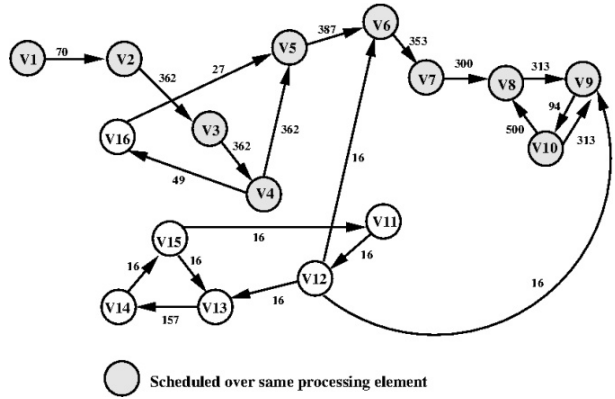

Figure 4. Scheduled IPs of VOPD

According to dynamic scheduling in place of IP core here cluster are map over processing element, if the number of cluster is greater than the available processing element, so in this case map cluster to the already mapped cluster i.e., combine the new cluster with the mapped cluster. This step is repeated until no cluster remains. In this way any number of clusters can be mapped over processing element.

Figure 5. Scheduled table of VOPD application

The goal of this mapping is to ensure that the threads that are expected to communicate frequently are assigned to neighboring processors as much as possible. The main objective of this mapping is to place a given data item into a node which is close to the nodes that access it. The last step of our approach determines the paths for data to travel in an energy efficient manner i.e. packet routing for that we incorporate here xyz routing algorithm.

## 6. BIO-INSPIRED OPTIMIZATION ALGORITHMS

In this section brief description of the implemented algorithms is discussed to provide optimal solution.

## 6.1 PARTICLE SWARM OPTIMIZATION (PSO)

Particle Swarm Optimization (PSO) is based on the movement and intelligence of swarms [23]. The fundamental idea behind PSO is the mechanism by which the birds in a flock (swarm) and the fishes in a school (swarm) cooperate while searching for food. Each member of the swarm called particle, represents a potential solution of the problem under consideration. Each particle in the swarm relies on its own experience as well as the experience of its best neighbor. Each particle has an associated fitness value. These particles move through search space with a specified velocity in search of optimal solution. Each particle maintains a memory which helps it in keeping the track of the best position it has achieved so far. This is called the particle0s personal best position *(pbest)* and the best position the swarm has achieved so far is called global best position *(gbest)*. After each iteration, the *pbest* and *gbest* are updated for each particle if a better or more dominating solution (in terms of fitness) is found. This process continues iteratively, until either the desired result is converged upon, or it's determined that an acceptable solution cannot be found within computational limits. In search of an optimal solution particles may trap in local optimal solution, therefore some heuristics can be used to help particles get out of local optimum. It is proved that PSO Algorithm discussed in [24] for IP mapping has worked much better than various proposed algorithms for routing and task mapping in NoC.

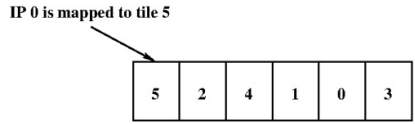

Figure 6. A particle representing a solution.

Basic PSO algorithm is implemented to find the optimal solution for IP mapping problem in NoC designs. Dimension or size (D) of the particle is set equal to the number of tiles in the NoC topology. Firstly, initial population with discrete values is generated having number of particles equal to the swarm size (S) and each particle is initialized with the initial velocity (V) and position (X). Particle for IP mapping is represented as shown in Figure 6, $n^{th}$ IP is mapped on $x_n$, where n is the index in particle. For IP mapping problem objective or fitness function (f) is to minimize *Total Communication Cost* described in Equation (1). Then these particles move into the search space in search of an optimal solution by updating their velocity and position towards its previous best solution (*P[s]*) and the global best solution (*P[best]*) using Equations (2) & (3).

$$v_i^{k+1} = wv_i^k + rand_1(0,c_1) \times (pbest_i - x_i^k) + rand_2(0,c_2) \times (gbest - x_i^k) \qquad (2)$$

and,

$$x_i^{k+1} = x_i^k + \lfloor v_i^{k+1} \rfloor \qquad (3)$$

where,
$\lfloor i \rfloor$ : gives floor value of i
$v_i^k$ : velocity of agent i at iteration k
w : weighting function
$c_1$ and $c_2$ : acceleration coefficients
rand : uniformly distributed random number between 0 and 1
$x_i^k$ : current position of agent i at iteration k

A modification is done in the position updating equation, the floor value of the updated velocity is added to the previous position for getting the new position of the particle, since the search space of IP mapping problem is discrete and the new velocity of particle may be real. For generating the integral position the floor value of velocity is taken. For performing experiments the values of various parameters for PSO are shown in Table 1(based on intensive experiments).

Table 1
PSO PARAMETERS AND THEIR VALUES

| Parameter | Value |
| --- | --- |
| $c_1$ | 1.2 |
| $c_2$ | 1.3 |
| w | 0.721348 |
| Swarm Size (S) | 200 |
| Dimension of the Search Space (D) | No. of tiles |
| Maximum No. of Simulations | 100 |
| Maximum No. of function evaluations in each simulation | 150000 |

## 7. EXPERIMENT RESULTS

In this section we present our experimental results derived from simulations of a combination of different real life applications and E3S benchmark. Our algorithm is compared with two other NoC-targeted mapping algorithms, and results are illustrated and analyzed in details.

Our proposed algorithm has been implemented in C++ and tested with different real life applications Video Object Plane Decoder (VOPD), Telecom, MMS (Multi-Media System), MWD (Multi-Window Display), randomly generated benchmarks using TGFF [25] and with E3S benchmark. These tasks are mapped over 3×3×3 3D NoC architecture. We have worked out with XYZ-routing algorithm which provides a deadlock free minimal routing technique. Results obtained are compared with results of various proposed approaches like Spiral, Crinkle which are proposed in literature [10].

Various parameter constraints that have been taken for performing the experiment are shown in Table 2. Number of benchmark over which the system has been tested having a 16 tasks *VOPD*, 16 task *MMS*, 12 tasks *Multi window display*, 16 tasks *Telecom*, 27 task *random* application and E3S(Embedded System Synthesis Benchmarks Suite) application which has 24 tasks *auto industrial*, 12 tasks *consumer*, 13 tasks *networking*, 5 tasks *office automation*.

By implementing existing and our proposed algorithms and comparing their result, when number of tasks are equal or less to number of processing element available we found that our algorithm gives optimal result. Testing our proposed algorithm with real life application, random benchmarks and E3S benchmark, we have noticed that our approach give better result (Figure 7) and provide improvement in the communication energy consumption (Figure 8). As discussed above, the average improvement in latency with various benchmarks is shown in Figure 9.

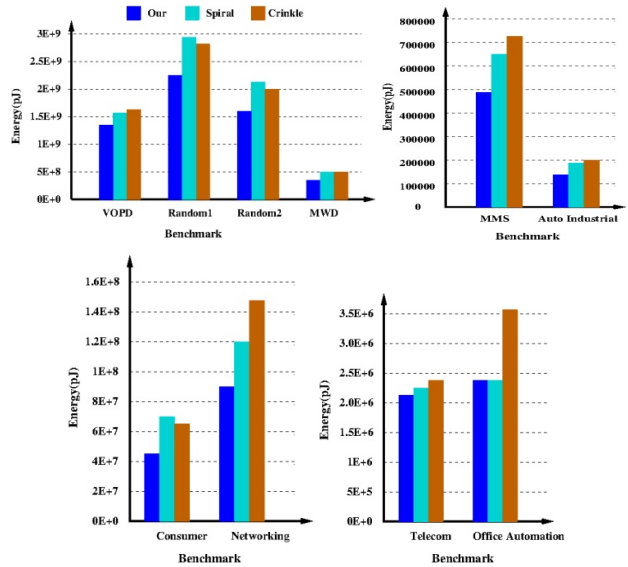

Figure 7. Energy Consumption with different benchmarks using various approaches.

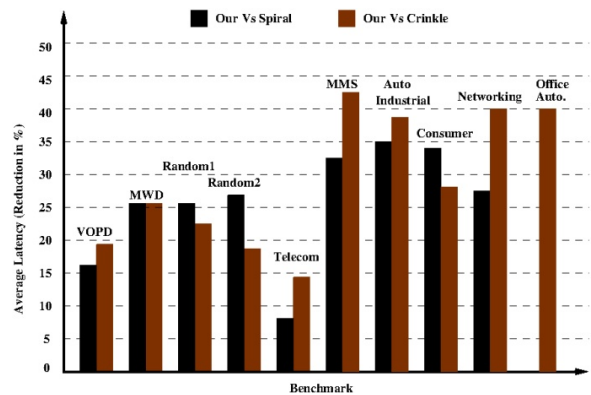

Figure 8. Energy Consumption reduction in % with various benchmarks.

Table 2.
BIT ENERGY VALUES FOR LINK AND SWITCH

| Link | Switch |
| --- | --- |
| 0.449pJ | 0.284pJ |

*Comparison of Dynamic & Cluster based Scheduling:*

Firstly, compare the energy consumption of a system with various real life standard benchmarks and with E3s benchmark shown in Figure 10 and results shows the improvement in the energy consumption of a system. Figure 12 shows the reduction in percentage form when we compare the above discussed proposed algorithm and when we evaluated the average reduction in energy consumption of a system is to be 49% with various benchmark.

Then above formulated average latency calculation equation is used to calculate the average latency with standards benchmark and E3S benchmark and comparison result is shown in Figure 11 which shows the improvement in the average delay time for transferring of a single bit. More energy is dissipated if transfer of single bit takes more time to reach the destination.

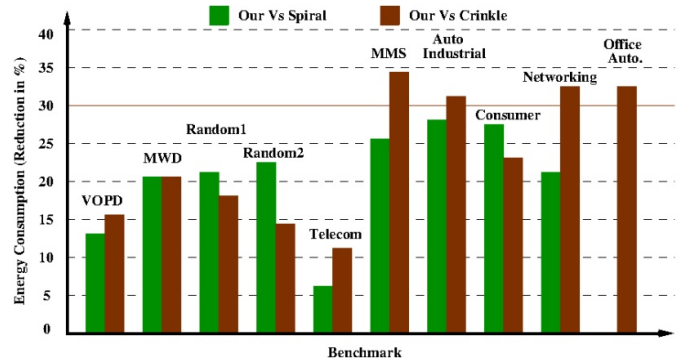

Figure 9. Average Latency reduction in % with various benchmarks

Figure 13 shows the reduction in percentage form when we compare the above discussed algorithm and we evaluated the overall reduction in average delay time of a system is to be 34% with various benchmarks, so according to the result shown this will definitely improve the performance of the system in terms of communication energy consumption and latency.

Figure 14 shows an improvement in the communication cost between source and destination and the communication cost is evaluated on the basis of number of bits transferred between source and destination and the number of hops count which will give the total communication cost required for the system. On the evaluation of overall reduction in communication cost of a system is found to be 48% with various benchmarks.

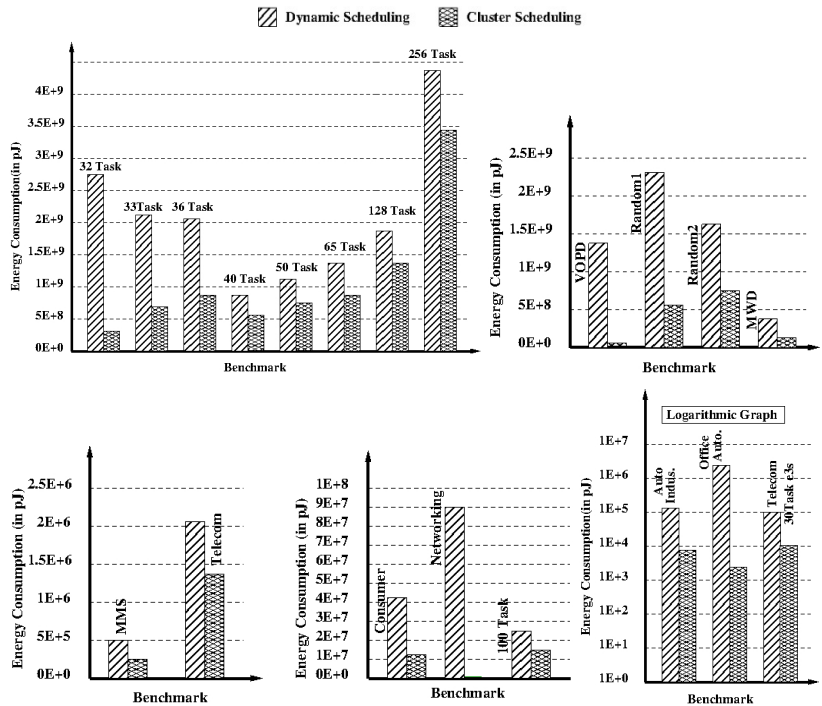

Figure 10. Energy Consumption (in pJ) with various benchmarks using dynamic and cluster scheduling

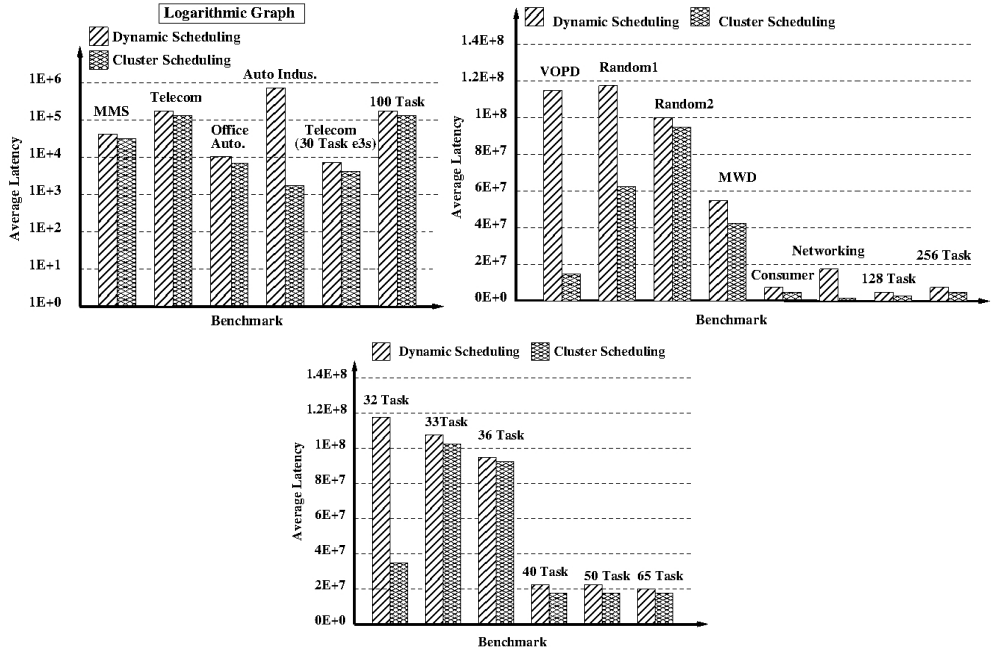

Figure 11: Average Latency with various benchmarks using dynamic and cluster scheduling

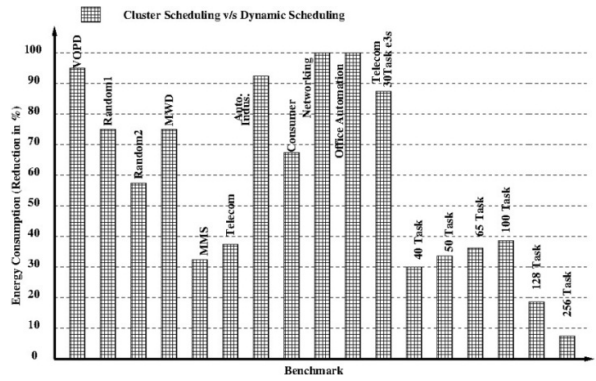

Figure 12. Energy Consumption (reduction in %) with various benchmarks for dynamic v/s cluster scheduling

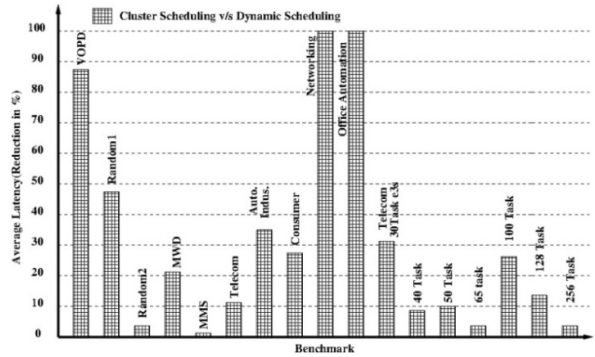

Figure 13. Average latency reduction with various benchmarks for dynamic v/s cluster scheduling

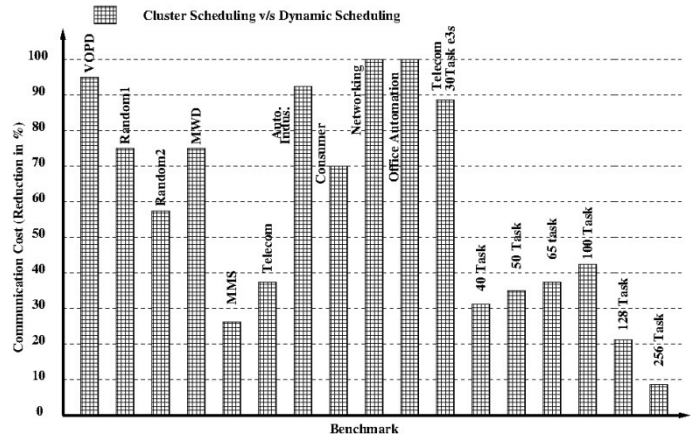

Figure 14. Communication Cost reduction with various benchmarks for dynamic v/s cluster scheduling

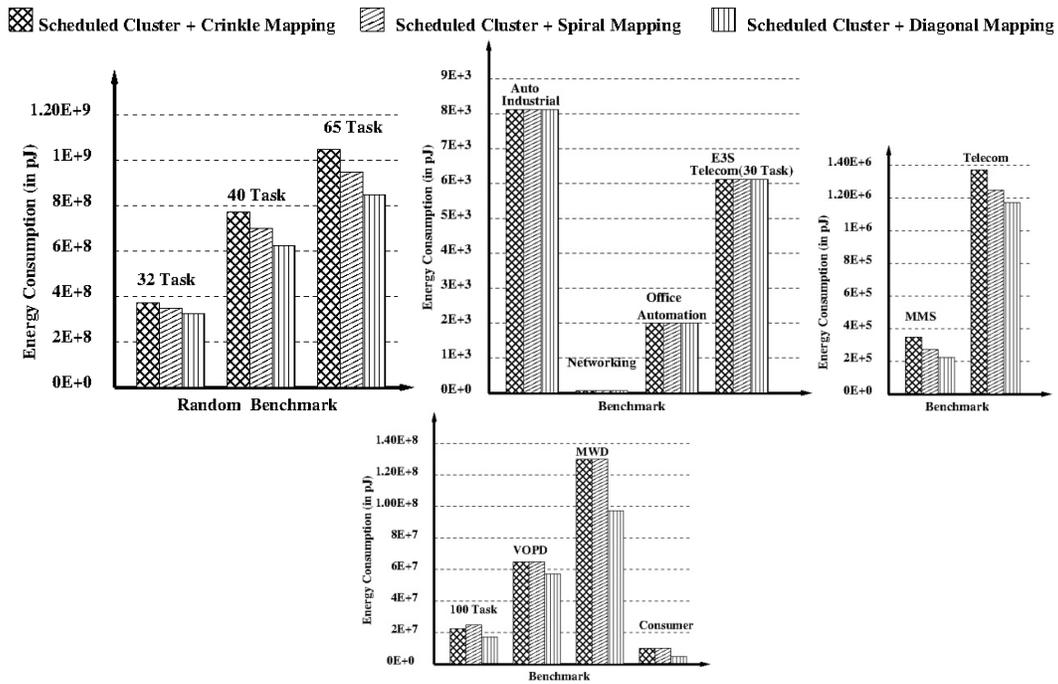

Figure 15. Energy Consumption reduction with various benchmarks for cluster based scheduling using crinkle, spiral and diagonal mapping styles.

*Comparison of Cluster Scheduling with DDmap, Spiral, Crinkle Mapping Style:*

The efficiency and efficacy of DDmap, Spiral, Crinkle algorithm has been evaluated by applying it to various real life benchmark multimedia applications, namely, Telecom, Multimedia system (MMS), Multi Window Display (MWD) and Video Object Plane Decoder (VOPD), random benchmark and the communication energy consumption is calculated for all the solutions generated and compared each result and found that diagonally mapping approach of cluster provide better result as compare to crinkle and spiral. Implementation of DDmap, Spiral and Crinkle algorithm are done in C++ language for three dimensional architecture and result are compared as shown in Figure 15. The implementation of spiral algorithm shown better result as compare to crinkle mapping algorithm but DDmap gives optimal result as compare to both. Through experimental analysis we can conclude that as the number of core increase the ratio of energy consumption between $3 \times 3 \times 3$ and $4 \times 4 \times 4$ architecture will increase. Main reason why spiral and crinkle gives worst energy as compare to DDmap is because of DDmap cover all the single link distance neighbor first then two hops distance neighbor and so on. But spiral start from center of mesh and continue mapping in spiral fashion while crinkle start with first PE of architecture and then continue mapping in a serial fashion that's why DDmap is better than both of them. Figure 15 experimental result shows that DDmap provides minimum energy consumption as well as low latency Figure 11.

*Comparison of Optimal Cluster Mapping using PSO*

In this section the results obtained by implementation of spiral and cluster mapping technique for problem of NoC Mapping are discussed [6]. It can be visualize from the graph that the applying of PSO for cluster mapping is giving better result for energy consumption as compare to spiral mapping show in Figure 16. This procedure incorporates a XYZ routing algorithm for packet routing. Similarly, PSO for cluster mapping is providing better result for energy consumption as compare to crinkle mapping of cluster shown in Figure 17. Clusters are combining on the basis of mapping style spiral and crinkle. But applying the PSO to these cluster mapping provide better results. The result shows that optimization using PSO provides better experimental result.

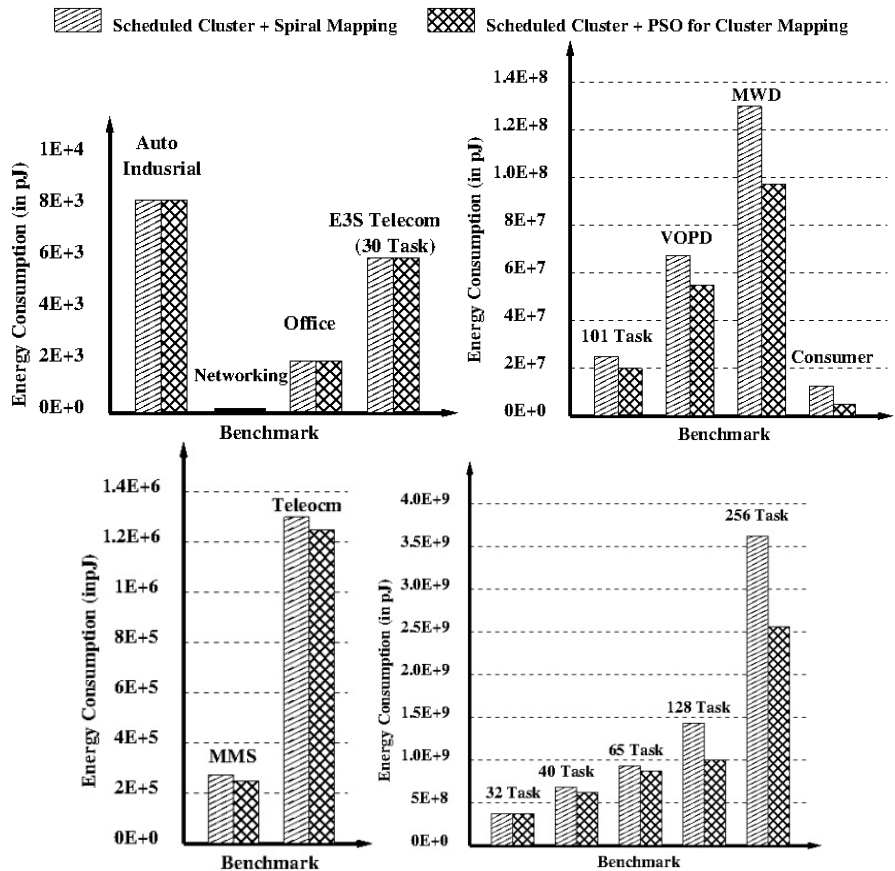

Figure 16. Energy Consumption reduction with various benchmarks for cluster based scheduling using spiral mapping and its optimization using PSO.

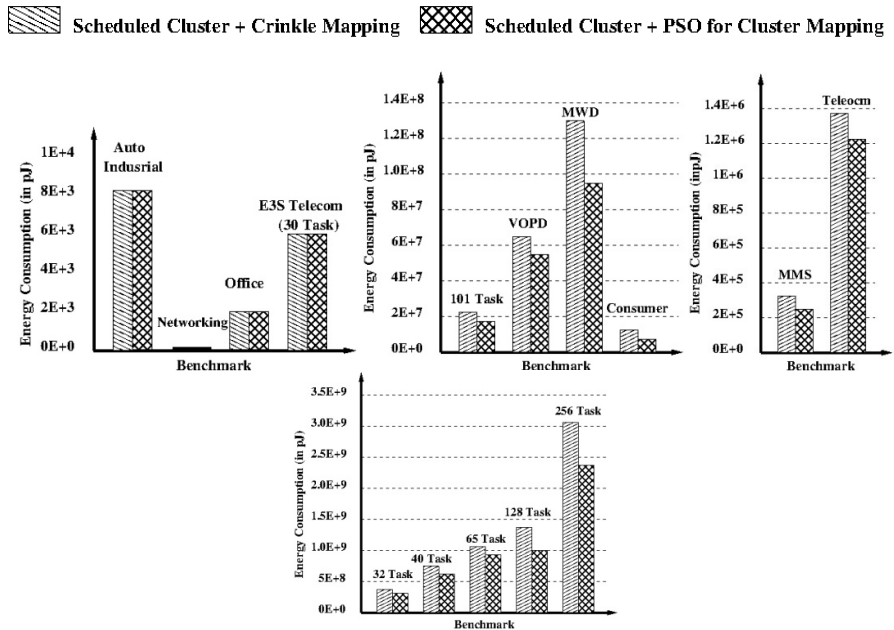

Figure 17. Energy Consumption reduction with various benchmarks for cluster based scheduling using crinkle mapping and its optimization using PSO.

## 8. CONCLUSION

In this paper, an energy efficient mapping approach for 3D NoC has been proposed. An efficient energy and performance aware scheduling algorithm is proposed which statically schedules both communication transactions and computation tasks onto homogeneous NoC architectures under real-time constraints. Although focused on the architectures interconnected by 3D mesh networks with XYZ routing schemes, proposed algorithm can be adapted to other regular architectures with different network topologies or different deterministic routing schemes. Number of real life applications is used to evaluate the performance of proposed algorithm and the results appear significantly better than contemporary NoC targeted mapping algorithms for 3D NoC. So, as discussed scheduling provide prominent solution for reduction in energy consumption and scheduling is an efficient solution for minimizing various parameter in NoC. So it is better to scheduled IP to system as librate to mapping, which has greater impact on performance of NoC. On analysis of experimental results, proposed algorithm provides improved results. Additionally, mapping the proposed approach to 3D NoC using DDmap, Crinkle and Spiral is carried out and observatory analysis provide prominent result with DDmap approach. Optimization of spiral and crinkle mapping is done using PSO which gives significant reduction in energy consumption.

## ACKNOWLEDGEMENT

I would like to thanks "*ABV- Indian Institute of Information Technology & Management Gwalior*" for providing me the excellent research oriented environment.

ISSN: 0975-3826

International Journal of Computer Science & Information Technology (IJCSIT) Vol 6, No 2, April 2014

**AUTHOR**

Vaibhav Jha has received his Master of Technology degree in specialization VLSI Design from Indian Institute of Information Technology and Management Gwalior in 2012. He has completed his Bachelors of Engineering degree in 2009. His academic research interest is in the area of Real time system, High Performance Computing, System on Chip, Distributed System, Computer Architecture, Databases, Networking and Communication.

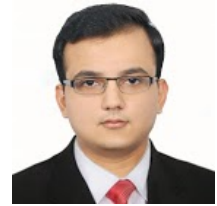